 \documentclass[final,5p,times,twocolumn]{elsarticle}

\usepackage{color}
\usepackage{graphicx}
\usepackage{epstopdf}
\usepackage{amsmath}
\usepackage[version = 3]{mhchem}
\usepackage{tabularx}
\usepackage{multirow}

\journal{Ultramicroscopy. Available online, 6 Dec 2018, doi:10.1016/j.ultramic.2018.12.004}
\tnotetext[a1]{\textsuperscript{\textcopyright } 2018. This manuscript version is made available under the CC-BY-NC-ND 4.0 license http://creativecommons.org/licenses/by-nc-nd/4.0/}

\begin{document}
\begin{frontmatter}
\title{The atomic lensing model: new opportunities for atom-by-atom metrology of heterogeneous nanomaterials}

\author[ua]{K.H.W. van den Bos}
\author[ua]{L. Janssens}
\author[ua]{A. De Backer}
\author[uo]{P. D. Nellist}
\author[ua]{S. Van Aert\corref{cor1}\fnref{phone}}
\cortext[cor1]{Corresponding author}
\ead{sandra.vanaert@uantwerpen.be}
\fntext[phone]{\emph{Phone}: +32 3 2653252}
\fntext[Phone]{Fax: +32 3 2653318}
\address[ua]{Electron Microscopy for Materials Science (EMAT), University of Antwerp, Groenenborgerlaan 171, 2020 Antwerp, Belgium}
\address[uo]{Department of Materials, University of Oxford, 16 Parks Road, Oxford, OX1 3PH, UK}

\begin{abstract}
The atomic lensing model has been proposed as a promising method facilitating atom-counting in heterogeneous nanocrystals \cite{vandenBos2016}. Here, image simulations will validate the model, which describes dynamical diffraction as a superposition of individual atoms focussing the incident electrons. It will be demonstrated that the model is reliable in the annular dark field regime for crystals having columns containing dozens of atoms. By using the principles of statistical detection theory, it will be shown that this model gives new opportunities for detecting compositional differences.
\end{abstract}

\begin{keyword}
High-resolution scanning transmission electron microscopy (HR STEM) \sep Quantitative STEM \sep Atomic lensing model \sep Data processing/image processing  \sep Statistical detection theory
\end{keyword}
\end{frontmatter}

\section{Introduction}
The exact three-dimensional (3D) atomic structure determines the properties of nanomaterials, making it essential to develop methods which are capable of characterising nanomaterials in 3D at the atomic scale. A popular characterisation technique is annular dark field (ADF) scanning transmission electron microscopy (STEM) because it provides structural images with sub-\aa ngstr{\"o}m resolution \cite{Batson2002, Erni2009}. The image intensities in these (partially) incoherent images are sensitive to both the thickness and the atomic number Z \cite{Pennycook1988, Nellist1999}, making them suitable to count the number of atoms in homogeneous nanocrystals \cite{LeBeau2010, VanAert2013}. These counting results enable one to retrieve the 3D atomic structure \cite{VanAert2011, Bals2011, Bals2012, Jones2014, DeBacker2017}. For heterogeneous nanocrystals, the dependency of the image intensities on the atomic number Z may be used to detect dopant atoms \cite{Ishikawa2014, Zhang2015} or determine the chemical concentration per atomic column \cite{Rosenauer2009, VanAert2009a, Martinez2014, Muller2016a}. The new atomic lensing model has been proposed to facilitate the expansion of the atom-counting technique from homogeneous to heterogeneous materials \cite{vandenBos2016}. In this manuscript, first the validity and applicability of the atomic lensing model is evaluated. Then, the possibilities and limitations for counting atoms and retrieving 3D structural information of heterogeneous nanocrystals from STEM images will be investigated.

In order to count the number of atoms, the so-called scattering cross-section has proven to be a successful image metric  \cite{VanAert2013, VanAert2011, Jones2014, VanAert2009a, Martinez2014, E2013}. It outperforms peak intensities and performs almost equally well as compared to a pixel-by-pixel analysis of the STEM image \cite{DeBacker2015}. Another advantage of this image metric is its robustness to probe parameters such as defocus, source coherence, convergence angle, and aberrations for well-separated columns \cite{E2013, Martinez2014a}. For counting atoms in homogeneous nanomaterials, one can use the monotonic increase of the scattering cross-sections with thickness. For heterogeneous materials, the atom-counting procedure is more complex since image intensities depend on thickness and the atomic number Z \cite{Ishikawa2014, Zhang2015, Martinez2014}. Furthermore, the 3D arrangement of atoms in a column also affects the scattering cross-section, as demonstrated in Fig. \ref{fig:VALalm}(a). To quantify these materials, a method is required that can recognise each possible 3D arrangement of atoms in a column. Therefore, the atomic lensing model is developed, which models dynamical diffraction as a superposition of individual atoms focussing the incident electrons \cite{vandenBos2016}. In this manner, scattering cross-sections of mixed columns can be predicted by using image simulations of pure columns. Then, the number of atoms can be estimated by comparing directly the predicted scattering cross-sections to calibrated experimentally measured values \cite{vandenBos2016, LeBeau2010, Rosenauer2009}. Hence, the atomic lensing model provides new opportunities to analyse heterogeneous nanomaterials at the atomic level.

By using the principles of statistical detection theory, the possibilities and limitations when analysing STEM images can be evaluated \cite{Dekker2013, Gonnissen2014, Gonnissen2016}. Here, discrete estimation problems, such as counting the number of atoms, are approached as a statistical hypothesis test \cite{DeBacker2015, Kay2009}. For heterogeneous materials, each hypothesis corresponds to a possible 3D arrangement of atoms in a column. Statistical detection theory enables one to calculate the probability to choose the wrong hypothesis, the so-called probability of error. In order to compute this probability of error, realistic simulations which describe the experimental images and knowledge about the statistics of the image pixel values are required. In electron microscopy, it is reasonable to assume that the image pixel values are ultimately Poisson distributed because of the unavoidable presence of electron counting noise \cite{Miedema1994}. Therefore, the resolving power with which different 3D arrangements of atoms in a column can be discriminated is fundamentally limited by the inherent presence of these statistical fluctuations. By minimising the probability to make an error as a function of the microscope parameters, the optimal experiment design can be derived. Here, only the optimal inner and outer collection angle of the annular STEM detector will be computed since the scattering cross-section is robust to probe parameters \cite{E2013, Martinez2014a}. It should be noted that the optimal detector collection angles do not necessarily correspond to the experimental settings leading to the best image contrast. The optimal detector settings will be used to investigate the possibilities and limitations for extracting 3D information of heterogeneous materials when using the atomic lensing model.

This manuscript starts by validating the use of the atomic lensing model and investigating under which experimental conditions the model may be applied (section \ref{sec:ALM}). By using the principles of statistical detection theory, the possibilities and limitations when characterising heterogeneous materials from STEM images are discussed in section \ref{sec:extract3D}. Finally, conclusions are drawn in section \ref{sec:conclusion}.

\section{The atomic lensing model}
\label{sec:ALM}
In order to count the number of atoms in heterogeneous materials, a method is needed that can recognise the scattering cross-section of any mixed atomic column. For a 20-atom-thick binary alloy having all possible ratios between both elements, there are already more than 2 million different 3D column configurations. Image simulations can provide this information, but the amount of computing time is tremendous. Therefore, this section investigates two different models that can predict the scattering cross-sections of mixed columns. First, in section \ref{sec:LinMod} a linear model neglecting electron channelling will be presented. Next, the atomic lensing model will be introduced in section \ref{sec:DeriALM}, which is non-linear and takes dynamical diffraction into account. In section \ref{sec:ValALM}, image simulations are used to validate the atomic lensing model and investigate the applicability of the model as a function of crystal thickness and detector settings. Furthermore, the advantages of the atomic lensing model as compared to the linear model neglecting channelling will be highlighted.

\subsection{The scattering cross-section for non-channelling conditions}\label{sec:LinMod}
By using the multislice approach, the electron specimen interaction is modelled by dividing a crystal into thin two-dimensional slices along the propagation direction of the electrons \cite{Cowley1957}. For calculating the electron wave function leaving the specimen, the wave function gets alternately transmitted through a slice and propagated to the next slice. In high angle ADF (HAADF) STEM imaging, it may be assumed that mainly thermal diffuse scattered (TDS) electrons will reach the annular detector because of its high collection angles \cite{Nellist2000}. In the multislice formalism, these electrons can be modelled by including the effect of thermally vibrating atoms. One option is to mimic the thermal vibrations by simulating for different thermal configurations of the specimen and averaging the resulting images. A computationally less intensive approximation has been suggested for STEM imaging by including an absorptive potential that accounts for TDS \cite{Pennycook1991, Ishizuka2002, Allen2003, Rosenauer2008a}. This approach is based on the original derivation by Yoshioka \cite{Yoshioka1957}, who added an extra energy term to the time-independent Schr{\"o}dinger equation in order to model the interaction energy between the electrons and the specimen. In the multislice formalism with absorptive potentials, the recorded intensity of the TDS electrons after a slice of thickness $d_z$ is given by \cite{Ishizuka2002}:

\begin{equation} \label{eq:I_TDS_slice}
\begin{split}
I_{TDS} \left( \mathbf{x}_0,z + d_z \right) = &\int \left| \psi \left( \mathbf{x}-\mathbf{x}_0, z \right) \right|^2 \times \\
&\left( 1 - \exp \left[-2 \sigma v_{TDS}^z \left( \mathbf{x} \right)\right] \right) d \mathbf{x}
\end{split}
\end{equation}

\noindent where $\mathbf{x} = (x,y)$ is the two-dimensional coordinate vector, $\psi(\mathbf{x}-\mathbf{x}_0,z)$ is the electron wave function at probe position $\mathbf{x}_0 = (x_0,y_0)$ and depth location $z$, $\sigma$ is the interaction parameter and $v_{TDS}^z(\mathbf{x})$ is the projected potential between $z$ and $z+d_z$ which describes the TDS electrons hitting the STEM detector. This potential is derived in Ishizuka \textit{et. al.} \cite{Ishizuka2002}. If the crystal is divided into $N$ slices of thickness $d_z$, the total recorded TDS intensity is given by:

\begin{equation} \label{eq:I_TDS}
I_{TDS}^{tot} ( \mathbf{x}_0) = \sum_{n=1}^{N} I_{TDS} \left( \mathbf{x}_0,n d_z \right). 
\end{equation}

When viewing a crystal along a major zone-axis orientation, the atoms are aligned in columns along the propagation direction of the incident electrons. Since the widths of both an atom potential and an aberration corrected electron probe are smaller than the distance between these atomic columns, the atoms in a column will only scatter electrons towards the STEM detector when the electron probe is in the vicinity of a column. Therefore, when scanning the electron probe over an atomic column, contributions to the recorded image intensity mainly results from this column \cite{Martinez2018}. Hence, the total HAADF STEM image intensity can be described as the superposition of the image intensities of the individual columns \cite{Martinez2014}. Since the scattering cross-section is equal to the total image intensity that is recorded when scanning the probe over an atomic column, one needs to consider only the potential of the atoms in this particular column. 

In the multislice approach, the crystal is commonly divided into thin slices containing only atoms at one particular depth location. In case of a single atomic column, every slice contains the projected potential of one atom. By assuming that the electron probe function does not change significantly within the crystal \cite{E2013}, the resulting recorded TDS intensity per slice is:

\begin{equation}\label{eq:I_TDS_sliceN}
I_{TDS} \left( \mathbf{x}_0,n d_z \right) = \int \left| \psi_0 \left( \mathbf{x}-\mathbf{x}_0 \right) \right|^2 \left( 1 - \exp \left[-2 \sigma v_{TDS}^n \left( \mathbf{x} \right)\right] \right) d \mathbf{x}
\end{equation}

\noindent where $\psi_0 \left( \mathbf{x}-\mathbf{x}_0 \right)$ is the incident electron probe function and $v_{TDS}^n(\mathbf{x})$ is the projected TDS potential of the $n$th atom in the column. 

For determining the scattering cross-section, the pixel values are integrated over all different probe positions around the column. In case of a single atomic column, the scattering cross-section for the atom in the $n$th slice is given by:

\begin{equation} \label{eq:SCSapproxInt}
\begin{split}
\Theta_{at}(n) = &\int \int \left| \psi_0 \left( \mathbf{x}-\mathbf{x}_0 \right) \right|^2 \times \\
&\left( 1 - \exp \left[-2 \sigma v_{TDS}^n \left( \mathbf{x} \right)\right] \right) d \mathbf{x} d \mathbf{x}_0.
\end{split}
\end{equation}

\begin{figure}[b!]
\centering
\includegraphics[width = 0.45\textwidth]{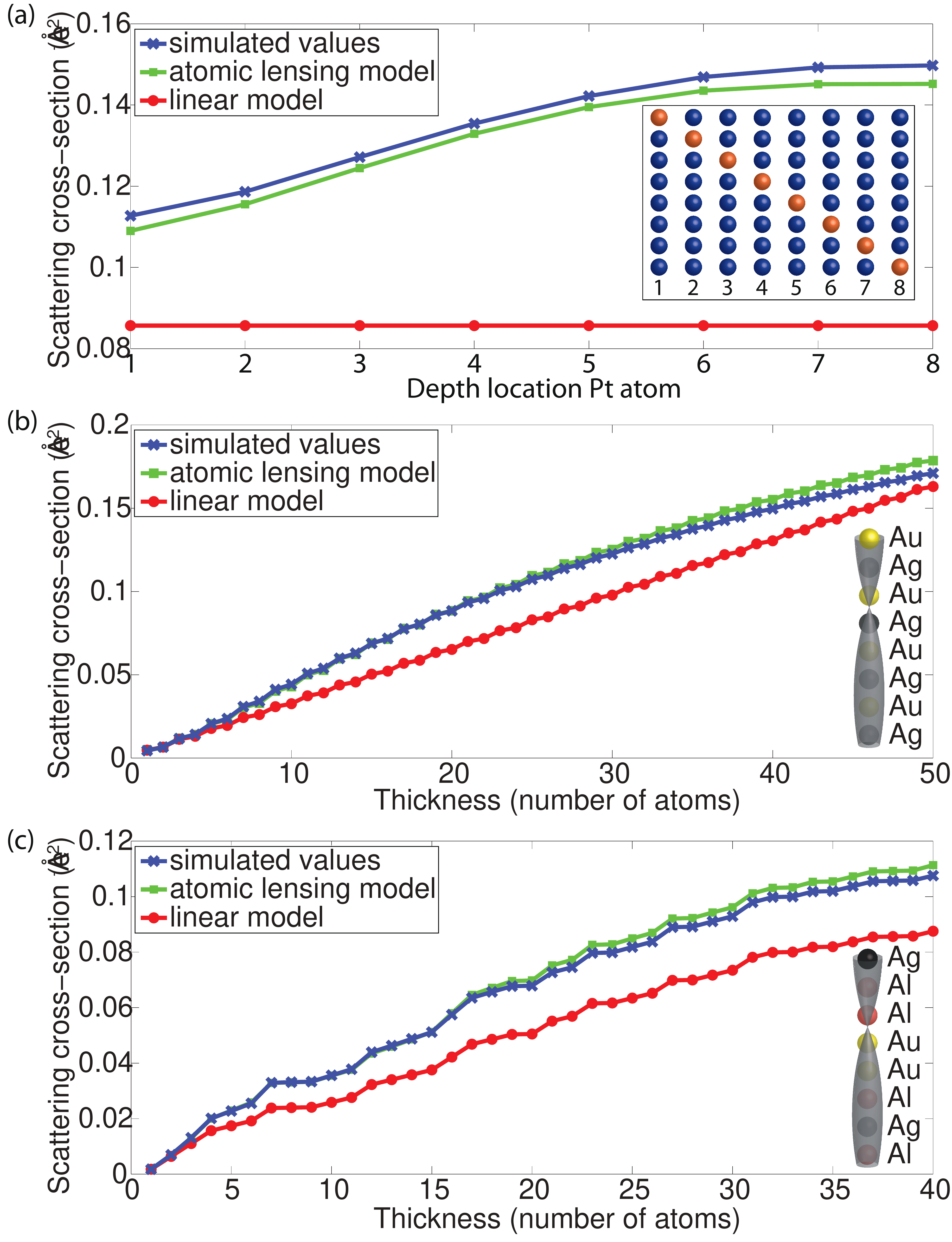}
\caption{Predicted scattering cross-sections by the atomic lensing model for different column configurations compared to image simulations and a linear model neglecting channelling. (a) An eight-atom-thick column of Co atoms containing 1 Pt atom at different depth locations. (b) A fcc Au-Ag crystal in [110] zone-axis orientation, with Au and Ag atoms alternating. (c) A column composed of Al, Ag, and Au in a random ordering. Image simulations are performed under the multislice approximation with absorptive potentials, with parameters as listed in Table \ref{tab:simPar}. Reprinted figure with permission from K.H.W. van den Bos, \textit{et. al.}, PRL 116 (2016), 246101. Copyright 2018 by the American Physical Society.}
\label{fig:VALalm}
\end{figure}

\noindent For quantitative comparisons between experimental images and image simulations, the total number of detected electrons is normalised relative to the number of incident electrons. As a result, the integral over the different probe positions equals:

\begin{equation} \label{eq:normProbe}
\int |\psi_0(\mathbf{x}-\mathbf{x}_0)|^2 d \mathbf{x} = \int |\psi_0(\mathbf{x}-\mathbf{x}_0)|^2 d \mathbf{x}_0 = 1.
\end{equation}

\noindent The scattering cross-section for the atom in the $n$th slice is then given by:

\begin{equation} \label{eq:SCSapprox}
\Theta_{at}(n) =  \int \left( 1 - \exp \left[-2 \sigma v_{TDS}^n \left( \mathbf{x} \right)\right] \right) d \mathbf{x}.
\end{equation}

\noindent This final expression indicates that the electron probe function does not influence the scattering cross-section, explaining the observed robustness to probe parameters such as defocus, source coherence, convergence angle, and aberrations for well-separated columns \cite{E2013, Martinez2014a}. In the derived expression, the scattering cross-section of an atom is independent of its depth location in the column. For example, the scattering cross-section of the $1$st and $n$th atom are equal in case of a pure, monotype, atomic column. Therefore, this expression suggests that a linear model, which adds the scattering cross-sections of the individual atoms, is suitable to predict the scattering cross-section of a mixed column composed of different atom types. However, based on image simulations, it has been found that this linear model, in which it is assumed that the probe wavefunction remains unchanged when it propagates through the crystal, leads to inaccurate estimates of the scattering cross-sections of mixed columns (see Fig. \ref{fig:VALalm}). Only for disordered crystals with moving atoms or small sample mistilt, a linear model has been found applicable \cite{MacArthur2015, Varambhia2016}. This demonstrates the need for a non-linear model in which dynamical diffraction is taken into account. This model will be derived in the next section \cite{vandenBos2016}. 

\begin{figure}[b!]
\centering
\includegraphics[width = 0.485\textwidth]{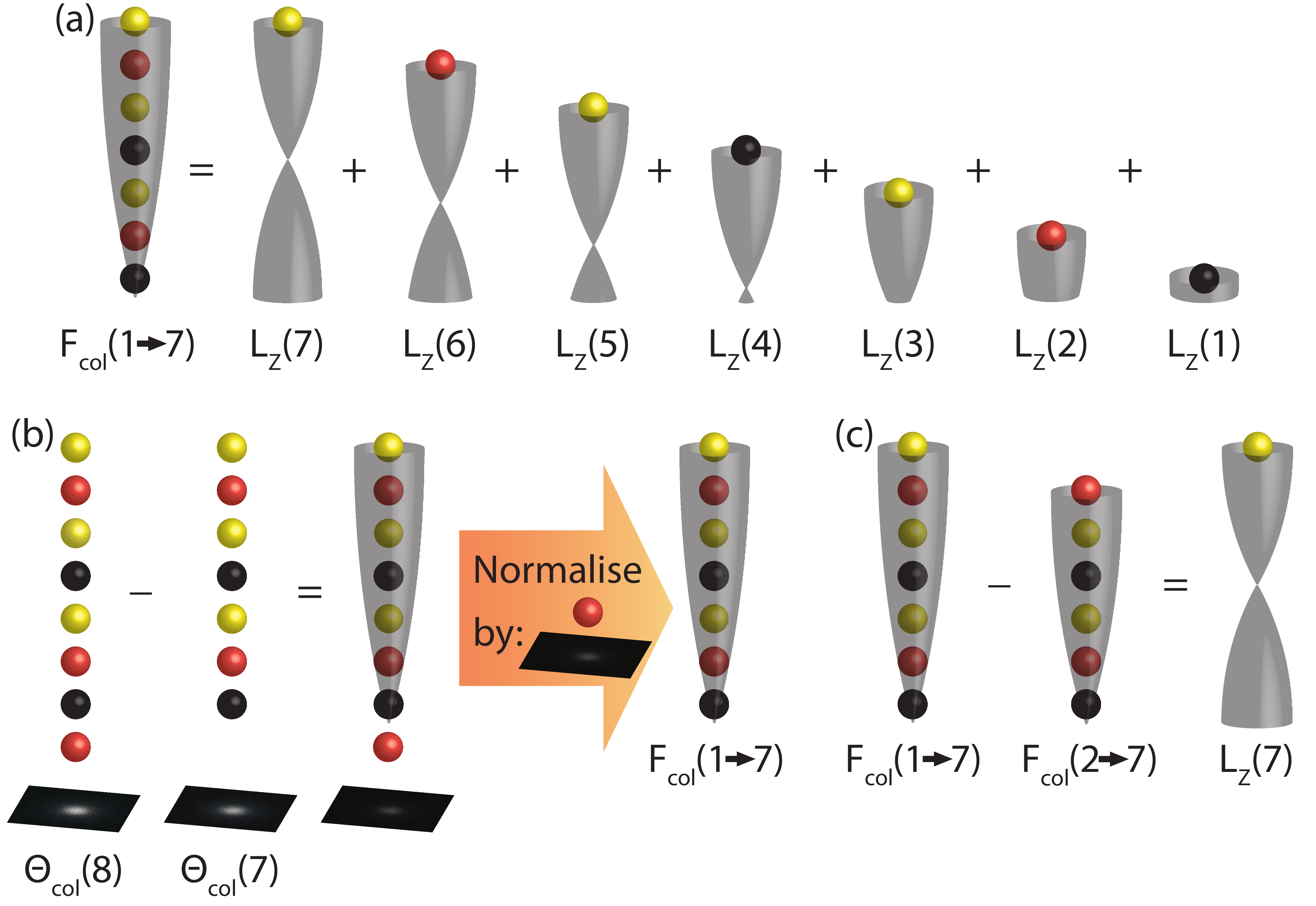}
\caption{The atomic lensing model visualised. (a) The lensing effect in an atomic column modelled as the superposition of individual atoms focussing electrons on the column. (b) The normalisation of the first derivative of the scattering cross-section with thickness by the scattering cross-section of a single free-standing atom gives the lensing effect of a column of atoms on an absolute scale. (c) Subtracting the lensing effect of different atomic columns gives the lensing effect of the individual atoms. Reprinted figure with permission from K.H.W. van den Bos, \textit{et. al.}, PRL 116 (2016), 246101. Copyright 2018 by the American Physical Society.}
\label{fig:ALM}
\end{figure}

\begin{figure*}[t!]
\centering
\includegraphics[width = 0.675\textwidth]{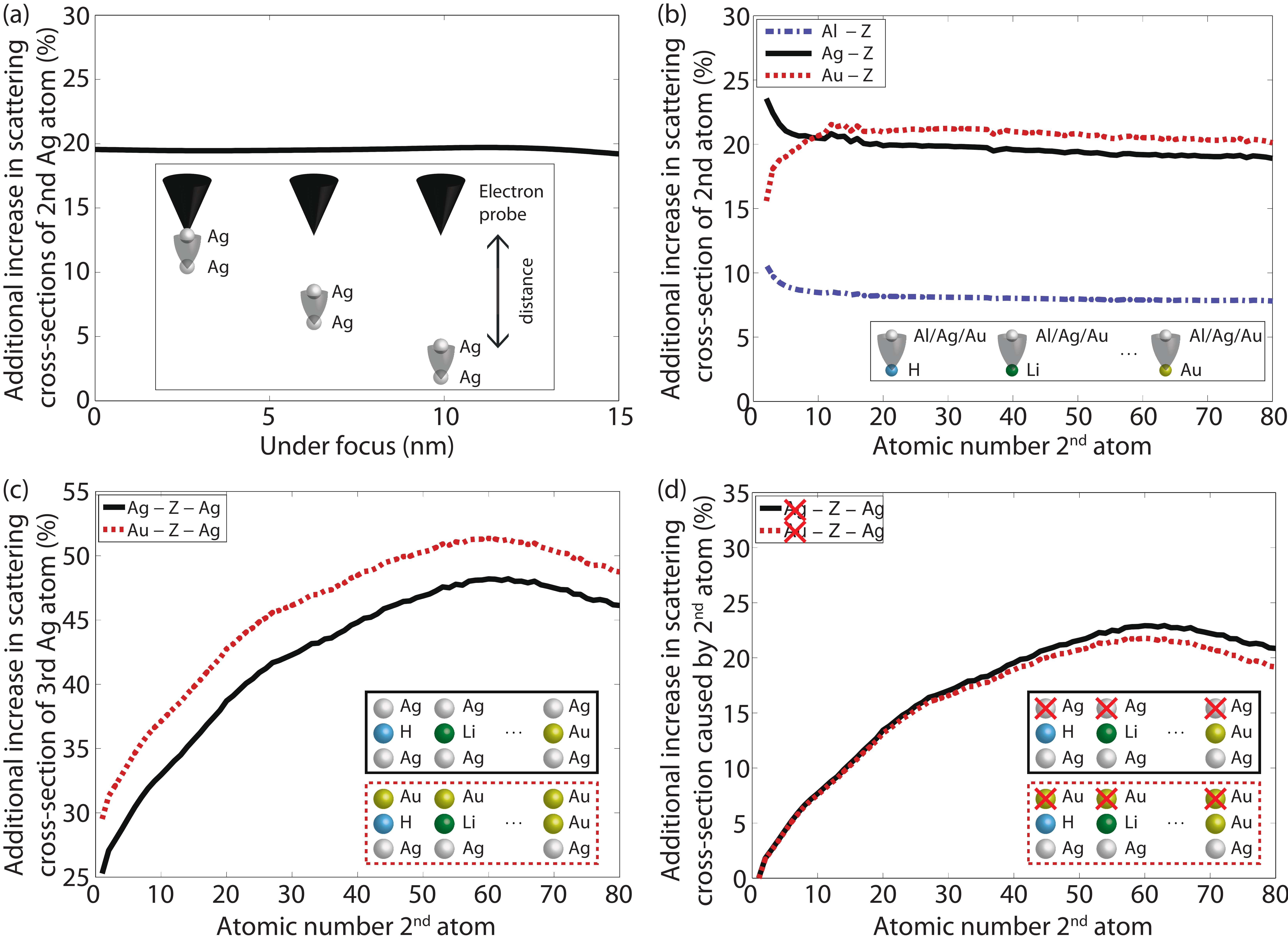}
\caption{Validation of the assumptions made in the atomic lensing model. The additional increase in the scattering cross-section of the $2$nd atom in a column as compared to the scattering cross-section of a single free-standing atom as a function of (a) the defocus value of the electron probe and (b) the atom type of the last ($2$nd) atom. (c) The additional increase in the scattering cross-section of the $3$rd atom in a column as compared to the scattering cross-section of a single free-standing atom. (d) Part of the additional increase caused by the $2$nd atom, obtained by subtracting the additional increase caused by the lensing effect of the $1$st atom from the results in (c). Image simulations are performed under the multislice approximation with absorptive potentials, with parameters as listed in Table \ref{tab:simPar}.}
\label{fig:ProofALM}
\end{figure*}

\subsection{Channelling effects on the scattering cross-section} \label{sec:DeriALM}
In the previous section, an expression for the scattering cross-section has been derived by assuming that the electron probe does not change in the crystal. In reality, the electron probe function is dynamically diffracted by the different atoms in the crystal and is not constant. To estimate scattering cross-sections of mixed columns in a reliable manner, the atomic lensing model has been proposed \cite{vandenBos2016}. Here, non-linearities are taken into account by considering dynamical diffraction. The starting point is the channelling theory, which assumes that each atom is a thin electrostatic lens that focusses the incident electrons \cite{VanDyck1996a, Cowley1997}. The atomic lensing model, describes this lensing effect as a superposition of individual atoms focussing the incident electrons (see Fig. \ref{fig:ALM}(a)). Here, it is assumed that the lensing effects of these individual atoms alter the electron probe function and cause a non-linear increase of the scattering cross-sections with thickness. In this case, the lensing effect of an atomic column as a whole is given by (illustrated in Fig. \ref{fig:ALM}(b)): 

\begin{equation}
F_{col}\left(1 \rightarrow n\right) = \frac{1}{\Theta_{col,Z\left(n+1\right)}\left(1\right)}\frac{d\Theta_{col}}{dn} = \frac{\Theta_{col}\left(n+1\right) - \Theta_{col}\left(n\right)}{\Theta_{col,Z\left(n+1\right)}\left(1\right)}
\label{eq:Fcol1ton}
\end{equation}

\noindent where $F_{col}\left(1 \rightarrow n\right)$ is the focussing effect of a column of $n$ atoms, with atoms located at the $1$st to $n$th position, and $\Theta_{col}\left(n\right)$ is the scattering cross-section of a column of $n$ atoms with $\Theta_{col}(0) = 0$. The increase in scattering cross-section (numerator) is normalised by $\Theta_{col,Z\left(n+1\right)}\left(1\right)$, corresponding to the scattering cross-section of a single free-standing atom with an atomic number equal to the $\left(n+1\right)$th atom to obtain the lensing effect of an atomic column. As shown in Fig. \ref{fig:ALM}(c), the lensing effect of an individual atom can then be determined from the superposition principle:

\begin{equation}
L_{Z}(n) = \frac{dF_{col}}{dn} = F_{col}\left(1 \rightarrow n\right) - F_{col}\left(2 \rightarrow n\right)
\label{eq:LZ}
\end{equation}

\noindent where $L_{Z}(n)$ is the lensing effect of the $1$st atom with atomic number $Z$ on the $(n+1)$th atom (equal to an atom's lensing effect over a distance of $n$ atoms). Similar as in optics, the lensing effect of an individual atom depends on the distance away from this atom. Therefore, the lensing effect of the $1$st atom on the $n$th atom is equal to the lensing effect of the $2$nd atom on the $(n + 1)$th atom (when the atomic numbers are the same). 

The model suggests that the lensing effect of an atom is independent of the surrounding atoms, enabling one to determine these values from image simulations of pure columns. The lensing effect of any mixed column, having in total $N$ atoms, can be determined by the superposition principle, allowing one to predict its scattering cross-section:

\begin{equation}
\Theta_{col}\left(N\right) = \Theta_{col}\left(N-1\right) +  \left(1+\sum_{n=1}^{N-1} L_{Z(n)}(N-n) \right)\Theta_{col,Z(N)}(1)
\end{equation}

\noindent where $Z(n)$ is the atomic number of the $n$th atom in a column. Since it is assumed that the lensing effects of the individual atoms, $L_{Z(n)}(N-n)$, alter the electron probe function, their contributions are added to a constant electron probe function. For predicting the scattering cross-section, this results in the addition of recorded intensities to the scattering cross-section of a single free-standing atom. Note that in the linear model, introduced in the previous section, $L_Z(n)=0$, resulting in an expression where the scattering cross-section of a column is computed by adding the scattering cross-sections of free-standing atoms.

\subsection{Predicting intensities of mixed columns} \label{sec:ValALM}
In the previous sections, a linear model and the atomic lensing model have been introduced to predict scattering cross-sections. While the linear model neglects channelling, the atomic lensing model assumes that the lensing effect present in the scattering cross-sections can be modelled as a superposition of individual atoms focussing the incident electrons. Both models assume that probe parameters hardly affect the scattering cross-sections. In order to validate these assumptions and to highlight the advantages of the atomic lensing model, image simulations of different column configurations have been used.

Image simulations have been performed under the multislice approximation with absorptive potentials, with parameters as listed in Table \ref{tab:simPar}. Debye-Waller factors have been obtained from Ref. \cite{Gao1999}. Mathematically, it has been shown in section \ref{sec:LinMod} that the electron probe function does not influence the scattering cross-sections when channelling is neglected. In order to verify whether the lensing effect in scattering cross-sections is affected by the electron probe function, the lensing effect of a Ag atom on a second Ag atom is computed as a function of probe defocus (a $1$st-order aberration) as is shown in Fig. \ref{fig:ProofALM}(a). Since the additional increase in the scattering cross-section caused by the lensing effect of the first Ag atom is constant for different defocus values, it seems that the electron probe function hardly influences the lensing effect. This suggestion is substantiated by the observed robustness of the scattering cross-section at different crystal thicknesses to probe parameters such as defocus, source coherence and aberrations \cite{E2013, Martinez2014a}. 

In the atomic lensing model, the additional increase of the scattering cross-section with thickness is normalised by the scattering cross-section of a single free-standing atom. It is therefore assumed that the lensing effect is independent of the atom type of the last atom. In order to validate this assumption, the additional increase in scattering cross-section as compared to a single free-standing atom is calculated for 2-atom-thick atomic columns using Eq. (\ref{eq:LZ}), where the atom type of the second atom is varied (see Fig. \ref{fig:ProofALM}(b)). This lensing effect as a function of the atom type of the second atom is calculated for a light (Al), heavier (Ag) and heavy (Au) atom. In all cases, the additional increase in the scattering cross-section is almost constant. Only for very light atoms changes are observed. These changes may be attributed to computational errors in the image simulations as the scattered signal of these light atoms is very weak in the HAADF regime for which the atomic lensing model is developed. Therefore, it can be concluded that the atom type of the last atom does not influence the lensing effect of any of the previous atoms.

For validating the assumption that the focussing effect of an atomic column can be written as a superposition of individual atoms focussing the incident electrons, image simulations of 3-atom-thick columns have been performed where the atom type of the $2$nd atom is varied and the atom type of the $1$st atom is either Ag or Au. By using Eq. (\ref{eq:Fcol1ton}), the additional increase in the scattering cross-section of the last Ag atom is calculated (Fig. \ref{fig:ProofALM}(c)). Furthermore, this equation is used to determine the additional increase caused by the lensing effect of the $2$nd atom, as shown in Fig. \ref{fig:ProofALM}(d). Here, the lensing effect of the $1$st Ag or Au atom is subtracted from the total lensing effect of the columns. Since the results in Fig. \ref{fig:ProofALM}(d) are overlapping, it can be concluded that indeed the lensing effect present in these 3-atom-thick atomic columns may be written as a superposition of individual atoms focussing the incident electrons.

\begin{table}[b]
\centering
\caption{Settings of the HAADF STEM multislice image simulations with absorptive potentials.}
\begin{tabular}{lll}
\hline
Parameter & Value \\
\hline
Acceleration voltage & $300~\mathrm{kV}$\\
Defocus &  $-83~$\AA\\
Cs condenser & $0.035~\mathrm{mm}$\\
Probe convergence semi-angle & $21.8~\mathrm{mrad}$\\
ADF inner collection angle & $90~\mathrm{mrad}$\\
ADF outer collection angle & $180~\mathrm{mrad}$\\
\hline
\end{tabular}
\label{tab:simPar}
\end{table}

\begin{figure}[t]
\centering
\includegraphics[width = 0.45\textwidth]{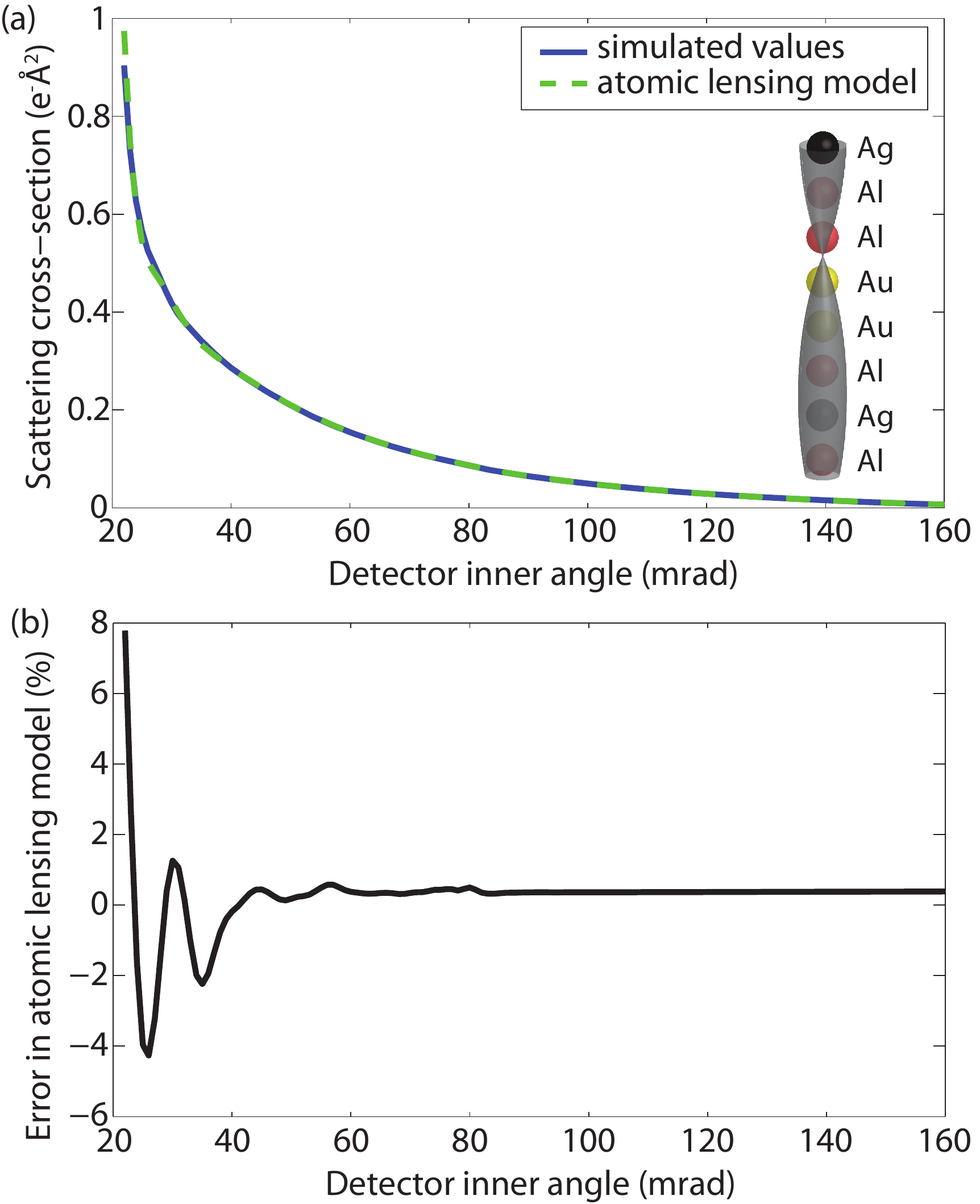}
\caption{The atomic lensing model evaluated as a function of detector innner angle. (a) The scattering cross-section of a 20-atom-thick column containing Al, Ag, and Au atoms in a random ordering. (b) The error in the predicted values by the atomic lensing model as compared to the simulated values. Image simulations are performed under the multislice approximation with absorptive potentials, with parameters as listed in Table \ref{tab:simPar}. Only the detector inner angle is changed, the outer angle is kept constant at 180 mrad.}
\label{fig:ALMdet}
\end{figure}

In reality, most atomic columns in nanomaterials contain more than 3 atoms. Therefore, comparisons with image simulations of different column configurations were made in Ref. \cite{vandenBos2016}, as shown in Fig. \ref{fig:VALalm}. The results indicated that the model preserves the dependency on the column configuration (see Fig. \ref{fig:VALalm}(a)) and is not restricted to a specific column ordering or maximum number of atom types, see Figs. \ref{fig:VALalm}(b) and \ref{fig:VALalm}(c). This is in contrast to the linear model, which shows large deviations and is therefore not suitable to predict the scattering cross-sections of mixed columns. Up to a thickness of about 25 atoms (10 nm), there is an excellent match between the atomic lensing model and the image simulations. Beyond this thickness still a decent match is found, indicating that the underlying superposition model holds approximately. Here, deviations are most likely caused by neighbouring atoms that may influence the lensing effect of a specific atom, both in-plane as well as out-off-plane. Therefore, most accurate predictions are expected in the HAADF regime, where the signal is more incoherent and atoms can affect the lensing effect of each other to a lesser extent. In order to determine the exact detector regime for which the atomic lensing model is valid, the scattering cross-section of a 20-atom-thick column containing Al, Ag, and Au atoms in a random ordering is determined as a function of the detector inner angle, as shown in Fig. \ref{fig:ALMdet}. The results indicate that the predicted values by the atomic lensing model are most accurate for large inner angles, starting at approximately 50 mrad. Since the probe convergence semi-angle is 21.8 mrad, this suggest that the atomic lensing model is well suited to be used in the HAADF regime, where the detector inner angle is approximately 2-3 times larger than the probe convergence semi-angle. For smaller inner angles, the signal becomes more coherent and small errors arise in the predicted values. These errors are only a few percent, indicating that the atomic lensing model may still give decent estimates in the ADF regimes where the detector inner angle is approximately 1-2 times larger than the probe convergence semi-angle. It should, however, be noted that the few percent error in this region is as large as the dependence of the scattering cross-section on the 3D arrangement of atoms shown in Fig. \ref{fig:VALalm}(a). Therefore, when using the atomic lensing model, it is advised to use the HAADF regime when searching for exact 3D information.  

\begin{figure}[t!]
\centering
\includegraphics[width = 0.475\textwidth]{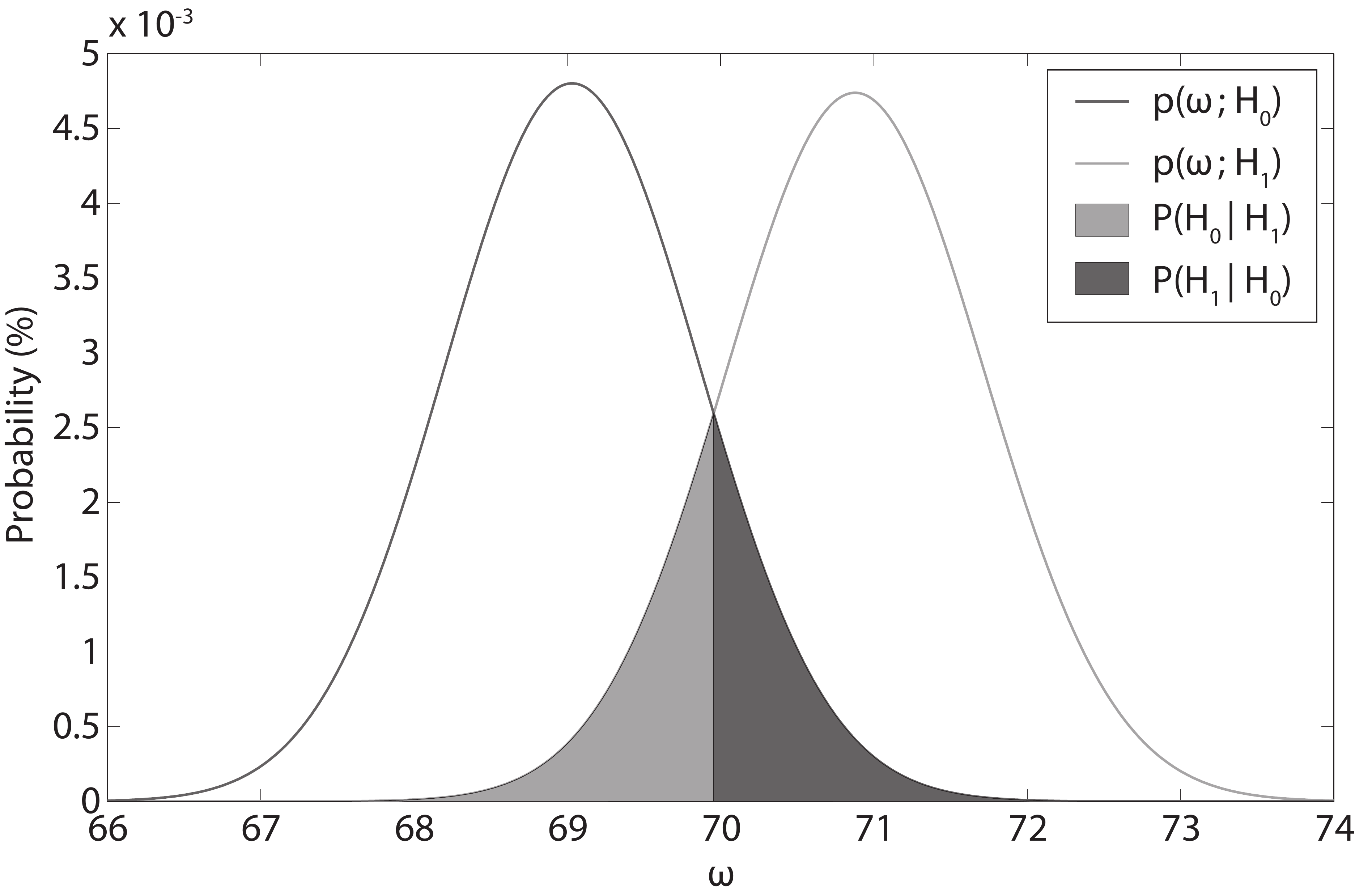}
\caption{Calculation of the probability of error for a binary hypothesis test on the scattering cross-section of a Pt and Au atom, corresponding to hypotheses $\mathcal{H}_0$ and $\mathcal{H}_1$, respectively. Since one chooses the hypothesis with the largest probability, the probability of error is given by the overlapping area. The probability functions, $p\left(\omega;\mathcal{H}_i\right)$, are determined from multislice STEM image simulations with parameters as summarised in Table \ref{tab:simPar} and collection angles of the STEM detector ranging from 60-300 mrad. $P\left(\mathcal{H}_i|\mathcal{H}_j\right)$ is the probability to decide $\mathcal{H}_i$ when $\mathcal{H}_j$ is true. The incident electron dose is 5 $\times$ 10$^5$ electrons per \AA$^2$. $\omega$ is equal to the scattering cross-section multiplied by the incident electron dose.}
\label{fig:PeExample}
\end{figure}

\section{Possibilities to locate the depth position of a Au impurity} \label{sec:extract3D}
In order to evaluate the possibilities and limitations for atom-counting and retrieving 3D information from single STEM images, the quantification of the depth location of a Au atom in a Ag column will be investigated. Here, statistical detection theory is used, which quantifies the probability to miscount the number of atoms or to misidentify the 3D arrangement of atoms in a column \cite{DeBacker2015, Dekker2013, Gonnissen2014, Gonnissen2016}. In this theory, problems are formulated as statistical hypothesis tests, where each hypothesis corresponds to a possible outcome; for example, two hypotheses that correspond to two possible atomic numbers that an atom may have. In STEM images, the pixel intensities may be considered as statistically independent electron counting results, which are Poisson distributed. Since the sum of Poisson distributed variables is again Poisson distributed, the probability function of the scattering cross-section can be calculated \cite{DeBacker2015}. In case of two hypotheses, the probability of error can therefore be visualised as the overlapping area of two probability functions as shown in Fig. \ref{fig:PeExample}. Note that when optimising an experiment, the overlap of the two probability functions should be minimised. In this section, the probability of error will be calculated based on De Backer \textit{et. al.} \cite{DeBacker2015}.

\begin{figure}[b]
\centering
\includegraphics[width = 0.5\textwidth]{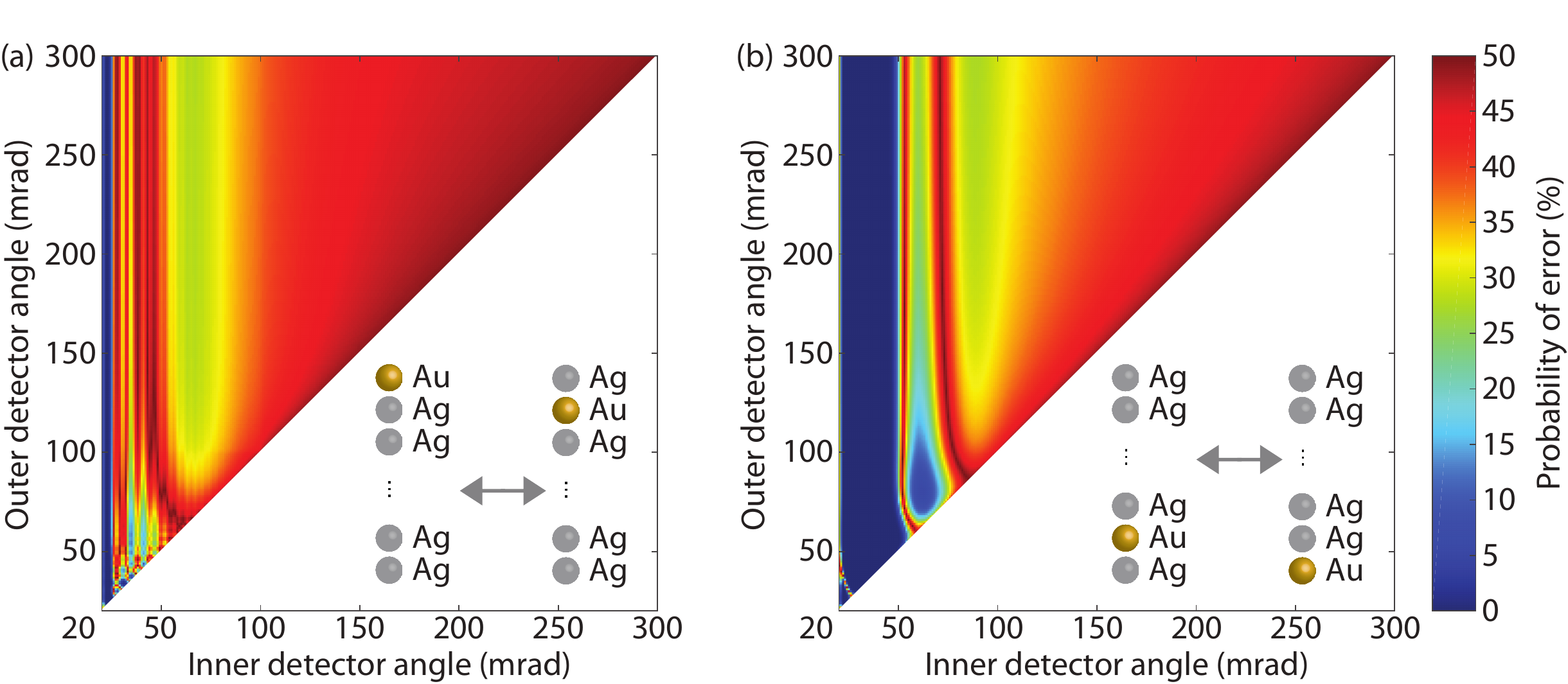}
\caption{Probability of error as a function of inner and outer detector angle to distinguish between 20-atom-thick Ag columns containing a Au atom at a different depth location, indicated in the figure. Results are obtained by using image simulations with parameters as listed in Table \ref{tab:simPar}. The optimal STEM detector angles correspond to a low probability of error. The incident electron dose is 5 $\times$ 10$^6$ electrons per \AA$^2$.}
\label{fig:DetAngles}
\end{figure}

\begin{figure*}[t!]
\centering
\includegraphics[width = 0.765\textwidth]{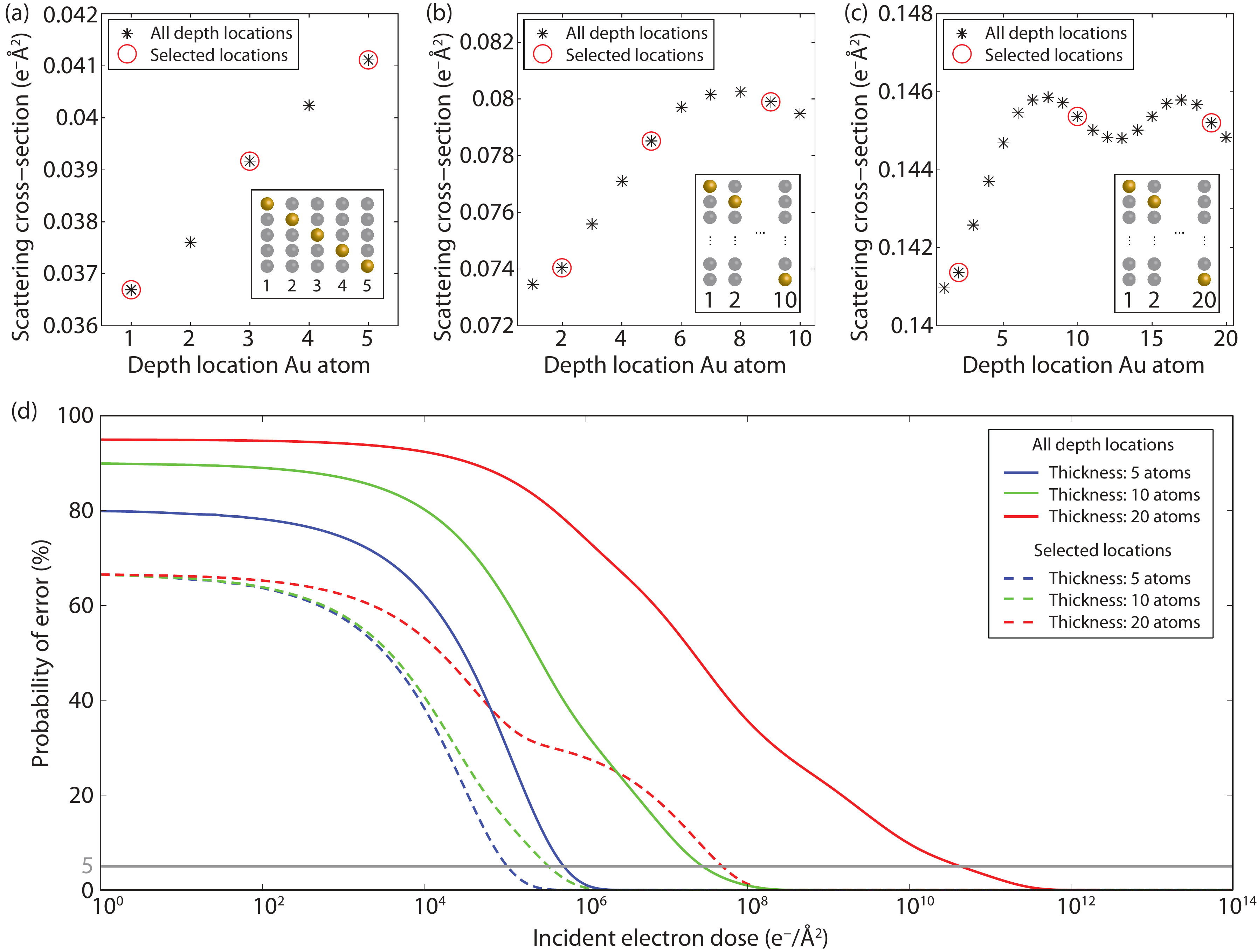}
\caption{The dependency of the scattering cross-sections on the depth location of a Au atom in a Ag column for a (a) 5-, (b) 10- and (c) 20-atom-thick column can be used to extract 3D information from a single viewing direction. (d) By using multiple hypothesis tests, the probability of error for identifying the exact depth location of the Au atom is computed as a function of the incident electron dose. Since a rough estimate on the depth location is sometimes sufficient, the reduction in required electron dose is computed when using only the 3 selected depth locations shown in (a),(b) and (c). The intersection of the results with the grey line indicates the minimally required electron dose needed to minimise the probability to misidentify a depth location to a maximum of 5\%.}
\label{fig:ReqDoseDepthLoc}
\end{figure*}

Before evaluating the possibilities when analysing STEM images of heterogeneous nanocrystals, optimal STEM detector settings are determined. Since the scattering cross-section is robust to probe parameters, such as defocus, source coherence, and lens aberrations \cite{E2013, Martinez2014a}, no optimisation is done for these parameters. For mixed columns, the number of hypotheses can be gigantic because it is equal to the number of possible 3D orderings of atoms in a column. Therefore, optimal STEM detector settings when quantifying the depth location of a Au atom in a Ag column are derived from binary hypothesis tests of 2 different scenarios, as shown in Fig. \ref{fig:DetAngles}. Here, 20-atom-thick Ag columns containing a Au atom close to the top and bottom are evaluated (see Figs. \ref{fig:DetAngles}(a) and \ref{fig:DetAngles}(b)). The results indicate that the LAADF regime is most suitable to measure the depth location. However, it can be seen that the suitable range of detector angles in this region is much more narrow for locating a Au atom close the top then for locating one close to the bottom. This can be explained by the coherent contributions to the signal that are present in the regime, making it questionable whether this regime is optimal for detecting each depth location. For atom-counting, it is shown that the coherent LAADF regime is no longer optimal when extending the analysis from binary test scenarios to multiple hypotheses \cite{DeBacker2015}. Other local optima can be found in the HAADF regime, where also the atomic lensing model predicts more accurate scattering cross-sections. Here, the optimal collection angles are ranging from either 60-300 and 90-300 mrad. As this second HAADF regime is only seen in the comparison where Au atoms are located at the bottom of the column, the most suitable STEM detector angles seem to range from 60-300 mrad. Therefore, these angles are used to investigate the possibility to find the depth location of a single Au atom in a column of Ag atoms from a single STEM image when knowing the thickness.

Ultimately, the dependence of the scattering cross-sections on the 3D ordering of atoms in a column can be used to extract 3D information from a single viewing direction. In order to investigate this possibility, scattering cross-sections have been computed for a 5-, 10- and 20-atom-thick Ag column (about 2, 4 and 8 nm thickness) containing a single Au impurity. Here, image simulations of a single pure Ag and Au atomic column have been performed with parameters as summarised in Table \ref{tab:simPar} and STEM detector angles ranging from 60-300 mrad, the suggested settings found before. The atomic lensing model has been used to determine the scattering cross-sections for the mixed columns containing 1 Au atom at different depth locations (see Figs. \ref{fig:ReqDoseDepthLoc}(a) - \ref{fig:ReqDoseDepthLoc}(c)). Since each depth location corresponds to one hypothesis, multiple hypothesis tests have been used which quantify the overlap of multiple probability functions as compared to the binary case shown in Fig. \ref{fig:PeExample} \cite{DeBacker2015}. In this manner, the probability of error as a function of the incident electron dose is determined for the 5-, 10- and 20-atom-thick columns, shown in Fig. \ref{fig:ReqDoseDepthLoc}(d). Here, it is assumed that only the depth location of the Au atom is unknown.

The results indicate that the probability to misidentify a depth location increases with increasing column thickness. Therefore, the electron dose required to identify each depth location is larger for thicker columns. In order to reduce the probability to misidentify a depth location to a maximum of 5\%, the minimally required electron doses are about $5 \times 10^5$, $3 \times 10^7$ and $4 \times 10^{10}$ electrons per \AA$^{2}$ for the 5-, 10- and 20-atom-thick columns, respectively. Especially, for the 20-atom-thick column, the required electron dose is too high to keep nanoparticles unharmed during an experiment. This can be explained by the depth dependency of the scattering cross-section, shown in Fig. \ref{fig:ReqDoseDepthLoc}(c). When moving the Au impurity from top to bottom, the scattering cross-section first increases but levels off when the Au atom becomes positioned in the centre of the column. Since differences between the scattering cross-sections of the different depth locations are extremely small, an unrealistically large incident electron dose would be required to distinguish between all these depth locations. For the thinner columns, this effect is less pronounced, lowering the minimally required electron dose. Therefore, these results suggest that the exact depth location of the Au impurity may only be found for beam-stable thin nanocrystals, with a thickness of about 4 nm at maximum. Reconstructing the exact 3D atom ordering for thicker columns from scattering cross-sections will be challenging when using a single image only.

Because of the low prospects for identifying the exact depth location, a second scenario is investigated where only a rough estimate on the depth location of the Au atom is given. Here, 3 different depth locations (top, centre and bottom) are chosen for each column thickness, shown in Figs. \ref{fig:ReqDoseDepthLoc}(a) - \ref{fig:ReqDoseDepthLoc}(c). In this manner, it is investigated whether the Au impurity is located close to the surface. The results in Fig. \ref{fig:ReqDoseDepthLoc}(d) indicate that for this problem a lower incident electron dose is required as compared to the problem where the exact depth location needs to be measured. For identifying the selected locations with a probability of error of at maximum 5\%, the required electron dosed are about $1\times10^5$, $3\times10^7$ and $5\times10^7$ electrons per \AA$^{2}$ for the 5-, 10- and 20-atom-thick columns, respectively. Especially for the 10- and 20-atom-thick column, the reduction in required electron dose is large. The results of the 20-atom-thick column shows a stepwise profile. This can be explained by the levelling off of the scattering cross-section with the depth location of the Au impurity, causing that 2 of the 3 selected locations have almost identical scattering cross-sections. Since the probability of error for this 20-atom-thick column first drops at an incident electron dose of about $10^5$ electrons per \AA$^{2}$, it indicates that at this dose a single Au atom located close to the top-surface can successfully be identified. Therefore, these results suggest that 3D information can be extracted from a single HAADF STEM image of a thin nanocrystal.

\section{Conclusions}\label{sec:conclusion}
In this manuscript, the validity and applicability of the atomic lensing model to predict scattering cross-sections of mixed columns has been investigated as a function of crystal thickness and STEM detector settings. This model facilitates atom-counting in heterogeneous nanocrystals as it allows one to identify the scattering cross-section of mixed columns having different 3D arrangements of atoms. The model takes non-linearities present in scattering cross-sections into account by describing dynamical diffraction as a superposition of individual atoms focussing the incident electrons. By using image simulations this assumption is validated and it is demonstrated that the lensing effect of an atom is independent of defocus. Furthermore, it is shown that up to a thickness of about 25 atoms (10 nm) the atomic lensing can reliably estimate the scattering cross-sections. Beyond this thickness the model can still decently predict values, in contrast to a linear model neglecting dynamical diffraction. Most reliable estimates are found in the HAADF regime, where the detector inner angle is approximately 2-3 times larger than the probe convergence semi-angle. For the smaller inner angles the errors are only a few percent, indicating that the atomic lensing model also gives decent estimates in the LAADF and MAADF regime.

Next, the possibilities and limitations for retrieving 3D structural information of heterogeneous nanomaterials have been investigated by using the principles of statistical detection theory. The results indicate that 3D information on the arrangement of atoms can be extracted from single HAADF STEM images. However, the exact 3D atomic ordering can only be retrieved from a single image for beam-stable thin nanocrystals, with a known thickness of about 4 nm at maximum. 

In the near future, the prospects may be exceeded as new direct electron detectors have become available which can recorded the full diffraction pattern per probe position \cite{Pennycook2015, Yang2015, Tate2016}. In this manner, it becomes possible to simultaneously record multiple STEM images with different detector settings. Combining different STEM images will enable one to extract more information from the same scan, which lowers the required electron dose and most likely makes quantitative STEM also accessible for beam-sensitive heterogeneous nanomaterials. In this future scenario, the atomic lensing model will be of even greater importance as these quantitative evaluations require image simulations for each detector setting.

\section{Acknowledgements} 
The authors acknowledge financial support from the Research Foundation Flanders (FWO, Belgium) through project fundings (G.0369.15N, G.0502.18N and WO.010.16N), and by personal grants to K.H.W.\ van den Bos and A.\ De Backer. This project has received funding from the European Research Council (ERC) under the European Unions Horizon 2020 research and innovation programme (grant agreement No. 770887).

\section*{References} 

\bibliographystyle{elsarticle-num}
\bibliography{Literature}

\end{document}